\documentclass[aps,twocolumn,groupedaddress,amsmath,amssymb]{revtex4-1}
\usepackage[title,titletoc,toc]{appendix}
\usepackage{graphicx}  
\usepackage{graphics}
\usepackage{dcolumn}   
\usepackage{multirow}
\usepackage{bm}        
\usepackage{verbatim}  
\usepackage{subfigure}
\usepackage{hyperref}
\usepackage{float}
\usepackage{epstopdf}
\usepackage{amssymb}
\usepackage{amsmath}
\hypersetup{
    colorlinks=true,
    linkcolor=blue,
    filecolor=magenta,      
    urlcolor=cyan,
}
\usepackage{booktabs}
\usepackage{xr-hyper}
\usepackage{stackengine}
\usepackage[USenglish]{babel}

\newcommand{\CAV}[1]{{\color{black} #1}}
\newcommand{\GG}[1]{{\color{black} #1}}

\begin{document}
    
    \title{Collective behavior of self-propelled rods with quorum sensing}
    
    \author{Clara Abaurrea Velasco}
    \affiliation{
	Theoretical Soft Matter and Biophysics, Institute of Complex Systems and Institute for Advanced Simulation, Forschungszentrum J\"ulich, D-52425 J\"ulich, Germany
    }
    \email{c.abaurrea@fz-juelich.de; g.gompper@fz-juelich.de; t.auth@fz-juelich.de}
    
    \author{Masoud Abkenar}
    \affiliation{
    Theoretical Soft Matter and Biophysics, Institute of Complex Systems and Institute for Advanced Simulation, Forschungszentrum J\"ulich, D-52425 J\"ulich, Germany
    }
    
    \author{Gerhard Gompper}
    \affiliation{
    Theoretical Soft Matter and Biophysics, Institute of Complex Systems and Institute for Advanced Simulation, Forschungszentrum J\"ulich, D-52425 J\"ulich, Germany
    }
    
    \author{Thorsten Auth}
    \affiliation{
    Theoretical Soft Matter and Biophysics, Institute of Complex Systems and Institute for Advanced Simulation, Forschungszentrum J\"ulich, D-52425 J\"ulich, Germany
    }
    
    \date{\today}
    
\begin{abstract}
Active agents -- like phoretic particles, bacteria, sperm, and cytoskeletal filaments in motility assays -- show a large variety
of motility-induced collective behaviors, such as aggregation, clustering and phase separation.
\GG{The behavior of dense suspensions of phoretic particles and of bacteria during biofilm formation is determined by 
two principle physical mechanisms: (i) volume exclusion (short-range steric repulsion) and (ii) quorum sensing (longer-range reduced propulsion
due to alteration of the local chemical environment).}
\CAV{To systematically characterize such systems, we study semi-penetrable} self-propelled rods in two dimensions, with a propulsion force that
decreases with increasing local rod density, by employing Brownian Dynamics simulations.
\CAV{Volume exclusion and quorum sensing both lead to phase separation, however, the structure and rod dynamics vastly differ.}
Quorum sensing enhances the polarity of the clusters, induces perpendicularity of rods
at the cluster borders, and enhances cluster formation. For systems, where the rods essentially become passive at high
densities, formation of asters and stripes is observed. \GG{Systems of rods with larger aspect ratios show more ordered structures compared 
to those with smaller aspect ratios, due to their stronger alignment}, with almost circular asters for strongly density-dependent propulsion force.
\GG{With increasing range of the quorum-sensing interaction, the local density decreases, asters become less stable, and polar hedgehog clusters and clusters with domains appear.}
Our results characterize structure formation and dynamics due to the competition of two qualitatively different \GG{interaction}
mechanisms, steric hindrance and \GG{quorum sensing}, which are both relevant for engineered phoretic
microswimmers as well as for bacteria \GG{in biofilm formation}.

\end{abstract}
    
\maketitle
    
\section{INTRODUCTION}
Many active systems in nature show collective behavior, ranging from sperm and bacteria \cite{goldstein_turbulence_2013, elgeti2015physics, winkler2015ecoli} to bird flocks, fish schools, and ant colonies \cite{gelblum2015ant, vicsek_collective_2012}. All these systems share a common characteristic: local alignment or jamming of 
neighboring agents gives rise to collective behavior.
This alignment can result from steric interaction between self-propelled elongated particles \cite{peruani_nonequilibrium_2006, baskaran_hydrodynamics_2008,abkenar_collective_2013}, but it can also emerge from other mechanisms, such as motility-induced clustering \cite{wysocki2014cooperative,stenhammar2014phase,abaurrea_rigid-ring_2017} and long-ranged, vision-like interactions \cite{turner_vision_2014, barberis_large-scale_2016}. In systems with steric interactions, the shape of the particles strongly influences the collective behavior. Disks and spheres, for instance, form round clusters \cite{dauchot_vibrated_2012, wysocki2014cooperative, volpe2011microswimmers,peruani_jams_2011}, while elongated objects, such as worm-like and rod-like particles, form elongated clusters and are often found in swarming phases \cite{isele2015self,abkenar_collective_2013, costanzo2014motility,yang_swarm_2010}. Motility assays with cytoskeletal filaments, such as actin filaments and microtubules, show clustering, swirling and wave-like patterns \cite{schaller2010polar, sumino_microtubules_2012}. Self-propelled particles (SPPs) can also be used to construct composite "complex objects" \cite{abaurrea_rigid-ring_2017,paoluzzi_vesicle_2016}, where the structure and dynamics of the self-propelled agents induce the motility of the composite particle.

\CAV{Quorum sensing is \GG{another} important mechanism to coordinate the behaviour of individual agents \GG{in dense suspensions, which results in a density-dependent}
propulsion force of SPPs. \GG{For bacteria in biofilm formation, such an interaction arises from a concentration field of signaling molecules, which
modifies the propulsion strength when the concentration exceeds a threshold \cite{wood_quorum_2006,sperandio_quorum_2002}.} 
\GG{
Genetically modified E.coli have been found to decrease their propulsion speed at high densities \cite{fu_bacteria_2012,liu_bacteria_2011}, which has been modeled by a density-dependent diffusion 
coefficient \cite{cates_bacteria_2010}.  This reduction of the swimming velocity helps biofilm formation. 
For phoretic SPPs, decreased propulsion at high densities due to a leveling of phoretic gradients} has been observed in 
simulations \cite{anderson_phoresis_1989,ripoll_nanoswimmer_2011,ripoll_flow_2013}. 
Phoretic propulsion mechanisms occur for particles in externally imposed gradients of solute concentration, electric potential, or temperature 
\cite{moran_phoretic_2017,ripoll_phoretic_2017}. When phoretic SPPs are close to each other, the gradients around the particles, and thus also the 
particle propulsion, decreases \cite{palacci_living_2013,theurkauff_dynamic_2012,moran_phoretic_2017,chiang_phoresis_2014}.
Finally, for fish schools, a reduced speed depending on both local density and polar order has been observed \cite{mishra_fish_2012}.}

SPPs accumulate where they move more slowly. Vice versa, they may also slow down at high densities, due to steric repulsion, biochemical signaling, or changes of the chemical environment. The positive feedback between accumulation-induced reduced propulsion and reduced propulsion-induced accumulation leads to motility-induced phase separation (MIPS) between a dense and a dilute fluid phase \cite{cates_mips_2015}.
\CAV{In a previous simulation study, a density-dependent reduced propulsion has been employed to mimic excluded-volume interactions in systems of point particles with
\GG{a Vicsek-type alignment rule} \cite{farrell_pattern_2012}. Bands, moving clumps, asters, and lanes have been reported.

In our simulations, we study for the first time the combination of steric interactions and \GG{quorum sensing, which are two physically distinct and independent mechanisms 
of density-dependent slowing down}, for systems of elongated SPPs. Our generic model for the collective behavior of bacteria and phoretic particles thus 
explicitely takes into account both shape and chemical signaling.}

\begin{figure}
	\centering
	\includegraphics[width=0.8\columnwidth]{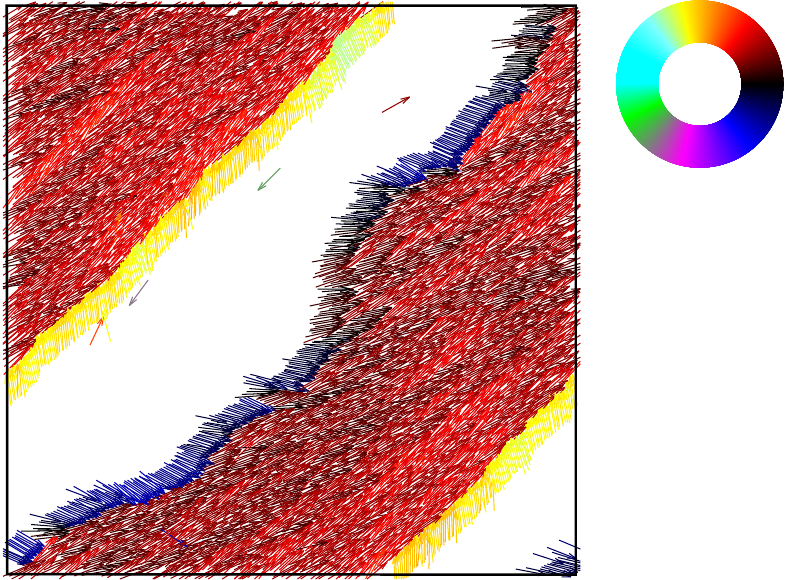}
	\caption{(Color online) Self-propelled rods with density-dependent reduced propulsion. Rods form polar clusters with perpendicular rods at the borders. System with aspect ratio $a/b=9$ $(n=18)$, $\rho L^2=12.8$, $E/k_\textrm{B}T=5$, $\textrm{Pe}=400$, \CAV{$\lambda=1$}, and $v_1=0.05$. Color wheel that indicates rod orientation.}
	\label{band}
\end{figure}    
    
We \CAV{simulate ensembles of self-propelled rods (SPRs) with a} propulsion force that decreases with increasing number of neighboring rods
in two spatial dimensions. The rods interact via a capped-repulsive potential that allows for crossing events, such that we effectively model a thin film with the computational costs of two-dimensional simulations \cite{abkenar_collective_2013}. The density-dependent propulsion force gives rise to a qualitatively different rod-rod alignment mechanism compared with the density-independent propulsion case. This leads to new phases that are not observed for SPRs with density-independent propulsion: \CAV{polar hedgehog clusters, asters, and polar clusters with perpendicular rods at the cluster borders}, see Fig.~\ref{band}. A density-dependent reduced propulsion force increases the polarity of the aggregates. Furthermore, increasing the range of the quorum sensing interaction destabilizes aster formation and promotes polar hedgehog clusters and clusters with domains.

\begin{figure}
        \centering
        \includegraphics[width=0.75\columnwidth]{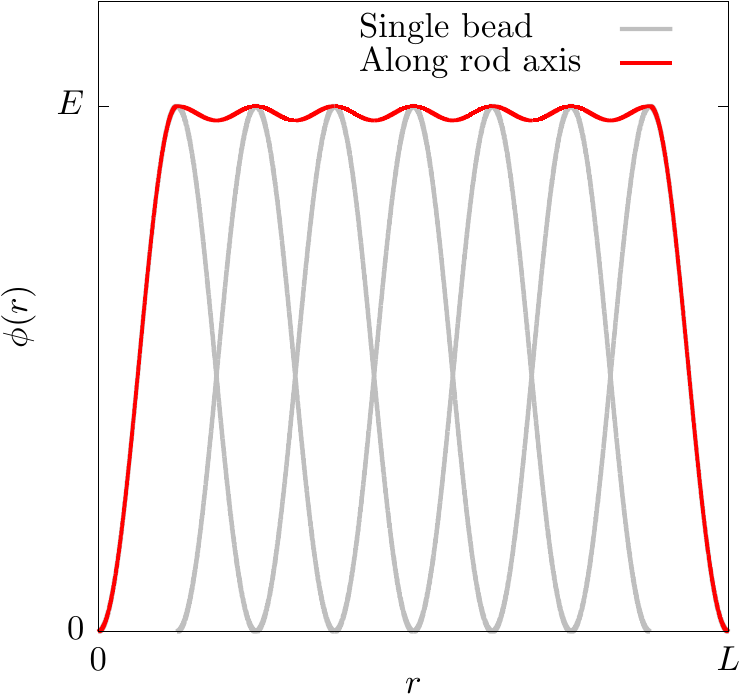}
        \caption{(Color online) Potential profile along a rod, $E$ is the rod energy barrier, and $L$ is the rod length. The gray curve represents the potential for the single beads, and the red curve is the sum of the contributions for all beads of the rod.} 
        \label{Rod_Potential}
\end{figure} 
    
Model, simulation technique, and numerical parameters are introduced in Sec.~\ref{sec:model}. Phase diagrams and a qualitative description of the collective behavior are presented in Sec.~\ref{sec:phase_diagrams}. In Sec.~\ref{sec:polar_order}, we quantify the effect of density-dependent reduced propulsion on the rod alignment and over-all polarity, using the polar order parameter. In Sec.~\ref{sec:density_cluster}, we quantify the effect of density-dependent reduced propulsion force on rod clustering, using rod density and cluster size distributions. In Sec.~\ref{sec:asters}, the perpendicular orientation of rods at cluster borders, as well as aster formation are highlighted. \GG{The effect of the range of the quorum-sensing interaction is investigated 
in Sec.~\ref{sec:quorum}.} In Sec.~\ref{sec:orient_correl}, we study rod dynamics using autocorrelation functions for rod orientation. Finally, Sec.~\ref{sec:conclusions} contains conclusions and outlook. Movies of the collective dynamics of the SPRs can be found in the electronic supporting information.
    
\section{MODEL AND SIMULATION TECHNIQUES}\label{sec:model}
We simulate SPRs with a density-dependent propulsion force using Brownian dynamics simulations in two dimensions. Our systems consist of $N$ rods of length $L$ in a system of size $L_x\,\times\,L_y$ with periodic boundary conditions. The rods are characterized by their center of mass positions $\mathbf{r}_{\textrm{r},i}$, their orientation angles $\theta_{\textrm{r},i}$ with respect to $x$ axis, their center-of-mass velocities $\mathbf{v}_{\textrm{r},i}$, and their angular velocities $\boldsymbol{\omega}_{\textrm{r},i}$ \cite{abkenar_collective_2013}.
    
\begin{figure}
    \centering
    \includegraphics[width=0.45\columnwidth]{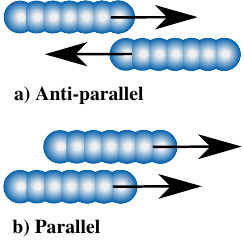} 
    \caption{(Color online) Schematic representation of rod-rod interaction and rod propulsion. \textbf{a)} Anti-parallel orientation between two rods, the angle between the rods is $\pi$. \textbf{b)} Parallel orientation between two rods, the angle between the rods is $0$.}
    \label{arrows}
\end{figure}
    
Rod-rod interactions are modeled using $n$ beads per rod, see Fig.~\ref{Rod_Potential}. Beads of neighboring rods interact via a separation-shifted Lennard-Jones potential (SSLJ) \cite{abkenar_collective_2013,fisher_sslj_1966}
\begin{equation}
\label{eqn:SSLJ}
\phi(r) = \begin{cases}
    4\epsilon\left[\left(\frac{\sigma^2}{\alpha^2 + r^2}\right)^{6} - \left(\frac{\sigma^2}{\alpha^2 + r^2}\right)^{3}\right] + \phi_0   & r \leq r_{\textrm{cut}}\\
    0 & r > r_{\textrm{cut}} 
    \end{cases}
\textrm{,}
\end{equation}
where $r$ is the distance between two beads, $\alpha$ characterizes the capping of the potential, and $\phi_{0}$ shifts the potential to avoid a discontinuity at $r=r_{\textrm{cut}}$. The length $\alpha=\sqrt{2^{1/3}\sigma^2-r_{\textrm{cut}}^2}$ is calculated by requiring the potential to vanish at the minimum of the SSLJ potential, $\sigma/r_{\textrm{cut}}=2.5$, hence the potential is purely repulsive. $E=\phi(0)-\phi(r_{\textrm{cut}})$ is the potential energy barrier. Once $E$ has been set to a certain value, we obtain $\epsilon=\alpha^{12}E/(\alpha^{12}-4\alpha^6\sigma^6+4\sigma^{12})$. 
With an effective bead radius $r_{\textrm{bead}}=r_{\textrm{cut}}/2$, and an effective rod thickness $r_{\textrm{cut}}$, the rod aspect ratio is $a/b=L/r_{\textrm{cut}}$. \CAV{The beads overlap a distance $r_\textrm{cut}$, see Fig.~\ref{Rod_Potential}, such that the effective friction for rod-rod interaction is small and no interlocking occurs \cite{abaurrea_rigid-ring_2017}.}
    
\begin{figure*}
    \centering
    \includegraphics[width=\textwidth]{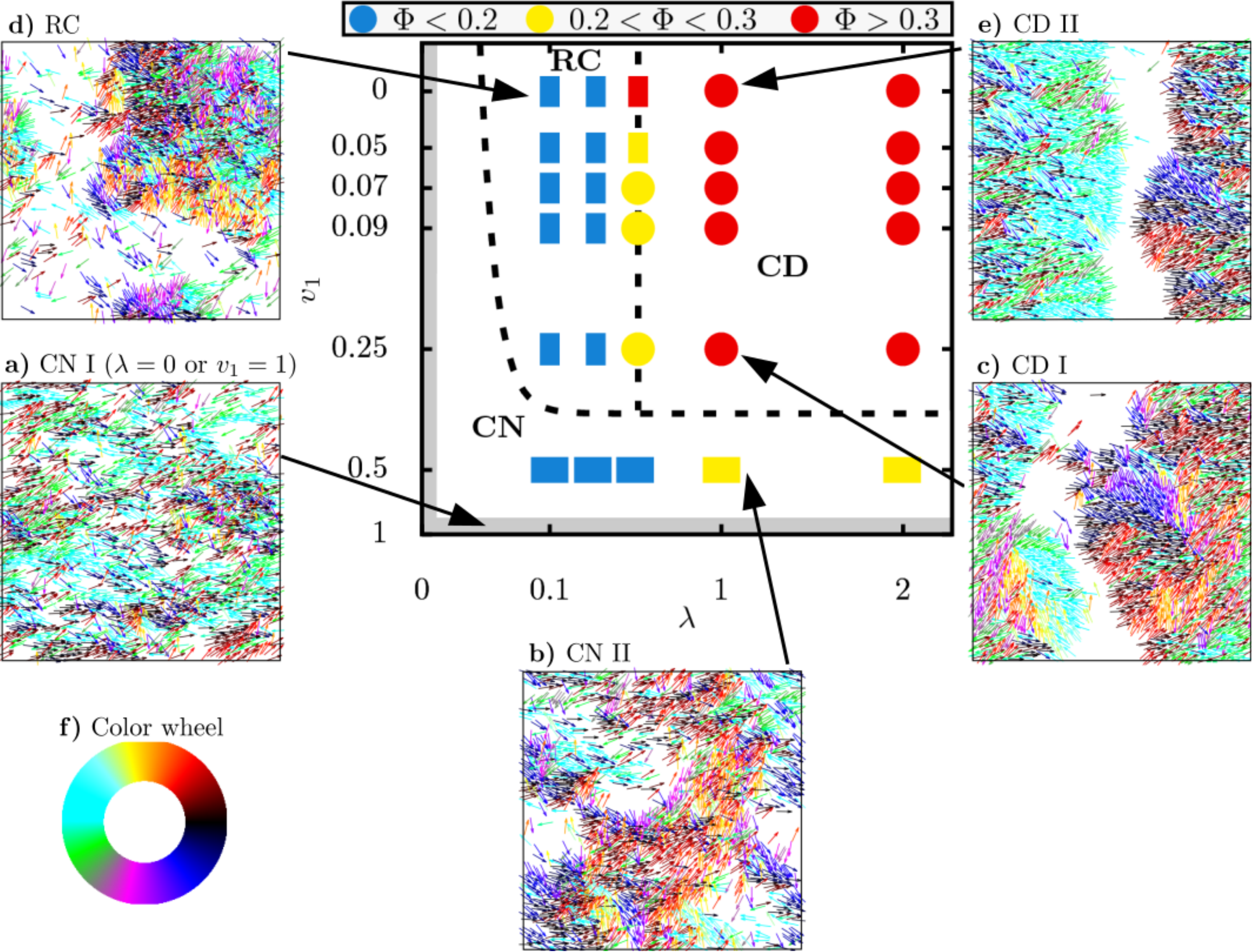}
    \caption{(Color online) Phase diagram for various $\lambda$ and $v_1$, and simulation snapshots of SPR systems with aspect ratio $a/b=4.5$ $(n=9)$, $\rho L^2=6.4$, $E/k_\textrm{B}T=5$ and $\textrm{Pe}=100$. \CAV{\textbf{a)} Clustered nematic phase with constant propulsion (CN I), i.e., $\lambda=0$. \textbf{b)} Clustered nematic phase with density-dependent slowing down (CN II), system with $\lambda=2$ and $v_1=0.5$. \textbf{c)} Clusters-with-domains phase (CD I), system with  $\lambda=2$ and $v_1=0.05$. \textbf{d)} Round clusters phase (RC), system with $\lambda=0.1$ and $v_1=0$. \textbf{e)} Clusters-with-domains phase with perpendicular rods at cluster borders (CD II), system with $\lambda=1$ and $v_1=0$.} \textbf{f)} Color wheel that indicates rod orientation. \CAV{The gray regions in the phase diagram indicate systems with density-independent propulsion, which correspond with $\lambda=0$ or $v_1=1$. In these regions the rods form a clustered nematic phase (CN).}   
    \CAV{In the phase diagram, squares represent the clustered nematic phase (CN), vertical rectangles represent round clusters (RC), and circles represent clusters with domains (CD).} The points are colored according to the value of the polar order parameter $\Phi$, see the legend. Note that the base propulsion weight $v_1$ appears in the vertical axis top to bottom.\CAV{ The penetrability coefficients range from $Q=0.2$ for density-independent systems ($\lambda=0$ or $v_1=1$) to $Q=0.001$ for systems with $\lambda=2$ and $v_1=0$.} Selected movies are presented in the Supplemental Material.}
    \label{snapshots_dens_l9} 
\end{figure*}
    
In our simulations, the rod velocity is decomposed into parallel and perpendicular components for the center-of-mass velocity, $\mathbf{v}_\textrm{r}=\mathbf{v}_{\textrm{r},\parallel}+\mathbf{v}_{\textrm{r},\perp}$, and the angular velocity $\boldsymbol{\omega}_{\text{r}}$,
\begin{equation*}
    \mathbf{v}_{\text{r}_i\parallel} = \frac{1}{\gamma_{\text{r}\parallel}}\left(\sum_{j\neq i}^{N} \mathbf{F}_{\text{r}_{i,j}\parallel}+\xi_{\textrm{r}\parallel}\mathbf{e}_{\parallel} + \mathbf{F}_{\text{p}}  \right)
\end{equation*}
    
\begin{equation*}
    \mathbf{v}_{r_i\perp} = \frac{1}{\gamma_{\text{r}\perp}}\left(\sum_{j\neq i}^{N} \mathbf{F}_{\text{r}_{i,j}\perp}+\xi_{\textrm{r}\perp}\mathbf{e}_{\perp}\right)
\end{equation*}
    
\begin{equation}
    \boldsymbol{\omega}_{\text{r}_i} = \frac{1}{\gamma_{\textrm{r}\theta}}\left(\sum_{j\neq i}^{N} \mathbf{M}_{\text{r}_{i,j}}+\xi_{\textrm{r}\theta}\mathbf{e}_{\theta}\right) \textrm{.}
\end{equation}
Here, $\mathbf{e}_{\parallel}$ and $\mathbf{e}_{\perp}$ are unit vectors that are parallel and perpendicular to the rod axis, respectively, and $\mathbf{e}_{\theta}$ is oriented normal to the plane of rod motion.
$\mathbf{F}_\text{p}$ is the propulsion force, $\mathbf{F}_{r_{i,j}}$ and $\mathbf{M}_{r_{i,j}}$ are force and torque from the interaction of rod $j$ and rod $i$, respectively. The rod friction coefficients, $\gamma_{\text{r}\parallel}=\gamma_0L$, $\gamma_{\text{r}\perp}=2\gamma_{\parallel}$ and $\gamma_{\text{r}\theta}=\gamma_{\parallel}L^2/6$, are obtained from hydrodynamic calculations for the rod in the slim-body approximation. 
The random noises $\xi_{\textrm{r}\parallel}$, $\xi_{\textrm{r}\perp}$, and $\xi_{\text{r}\theta}$ are drawn from Gaussian distributions with variances $\sigma^2=2k_\textrm{B}T\gamma_0/\Delta t$ \cite{abkenar_collective_2013, loewen_spherocylinders_1994}, where $\Delta t$ is the time step used in the simulations. Thus, ensuring that the fluctuation-dissipation theorem is fulfilled, at equilibrium.
    
There are three \CAV{energy scales} in our systems: the thermal energy $k_\textrm{B}T$, the propulsion strength $F_\text{p}L$, and the energy barrier due to rod-rod interactions $E$. Dimensionless ratios can be used in order to characterize the importance of the different contributions. The P\'eclet number \cite{abkenar_collective_2013,abaurrea_rigid-ring_2017}
\begin{equation}
\textrm{Pe} = \frac{LF_\text{p}}{k_\textrm{B}T} 
\end{equation}
is the ratio of propulsion strength to noise. The dimensionless ratio that compares the product of propulsion strength with the rod repulsion energy barrier is the penetrability coefficient \cite{abkenar_collective_2013}
\vspace{-0.2cm}
\begin{equation}
\text{Q} = \frac{LF_\text{p}}{E} \textrm{.}
\end{equation}
We employ the density-dependent propulsion force \cite{farrell_pattern_2012}
\begin{equation}
\mathbf{F}_\text{p}=\mathbf{F}_0\left(v_0 e^{-\lambda m/n} + v_1 \right) \textrm{,}
\label{ddprop}
\end{equation}
\CAV{where $\mathbf{F}_0$ is the rod propulsion strength \GG{in the absence of slowing-down}, $v_0$ is the weight of the density-dependent propulsion force, $v_1$ is the weight of the base propulsion force, $m$ is the number of neighboring beads surrounding the rod, and
\begin{equation}
\lambda = \lambda_1\left(\frac{r_\textrm{cut}}{r_\textrm{int}}\right)^2 \textrm{,}
\label{eq_lambda}
\end{equation}
where $\lambda_1$ the base deceleration ratio.
For quorum sensing, $\lambda$ models the sensitivity to the concentration of signaling molecules. The interaction radius $r_{\rm int}$ can be interpreted as range of chemical signaling, determined by diffusion coefficient and degradation rates of the signaling molecules.} 
We choose $v_0+v_1=1$, such that for systems without neighbors $\mathbf{F}_\text{p}=\mathbf{F}_0$.
The number of neighboring beads, $m$, is calculated on a bead basis. 
For each bead of rod $i$, we calculate the number of beads of neighboring rods that are inside an area of \CAV{the interaction radius $r_\textrm{int}$}. We then sum the number of neighbors over the number of all beads of rod $i$ to obtain the total number of neighboring beads.
    
Although passive rods are apolar, the \CAV{finite rod thickness and the} rod-rod friction because of the discretization into beads \CAV{lead to a polar interaction \cite{abaurrea_rigid-ring_2017,abkenar_collective_2013}. A reduced propulsion force at high densities can be expected to enhance the polarity, because for} two rods in anti-parallel orientation the density-dependent deceleration lasts for $\tau_{\text{anti-parallel}}\approx L/(2\arrowvert \mathbf{v}_\parallel\arrowvert)$, while for two rods in parallel orientation the density-dependent deceleration lasts much longer.
    
In the simulations, we employ \GG{dimensionless} units and parameters. \GG{Lengths are measured in units of the rod length $L$}, energies in units of $k_\textrm{B}T$, 
and times in units of $\tau_{0}=1/D_{\textrm{r}0}$, where $D_{\textrm{r}0}$ is the rod rotational diffusion coefficient.
\CAV{$1\leq r_\textrm{int}/r_\textrm{cut}\leq3$, unless explicitly stated otherwise, we use $r_\textrm{int}/r_\textrm{cut}=1$.}
The global rod density is ${\rho}_0=N/({L}_x\times {L}_y)$, where $N$ is the rod number.
The system size is ${L}_x={L}_y=16{L}$.
The systems that we have studied consist of rods with aspect ratios $a/b=4.5$ and $9$, $n=9\text{ and }18$ beads, respectively. 
\CAV{For our studies with $r_{\rm int}=r_{\rm cut}$, we packing fraction $\phi=1.4$, which corresponds to rod densities ${\rho}_0{L}^2=6.4$ for systems with $a/b=4.5$, and ${\rho}_0{L}^2=6.4\text{ and }12.8$ for systems with $a/b=9$. 
For our studies of different interaction radii, we use packing fraction $\phi=0.8$, which corresponds a rod density ${\rho}_0{L}^2=6.4$ for systems with $a/b=9$.}
Unless explicitly stated, rods with aspect ratio $a/b=9$ have densities of ${\rho}_0{L}^2=12.8$
We study systems with P\'eclet numbers $25\leq \textrm{Pe} \leq400$, rod energy barriers $1.5 \leq E/k_\textrm{B}T \leq 10$, base propulsion weights $0\leq v_1 \leq0.5$, and \CAV{base deceleration ratios $0.05\leq \lambda r^2_\textrm{int}/r^2_\textrm{cut} \leq2$. For the parameters shown here, we find penetrability coefficients $0.001\leq Q\leq0.8$, which correspond to impenetrable rods.} Rod positions and orientations are initialized randomly.
    
\begin{figure*}
    \centering
    \includegraphics[width=\textwidth]{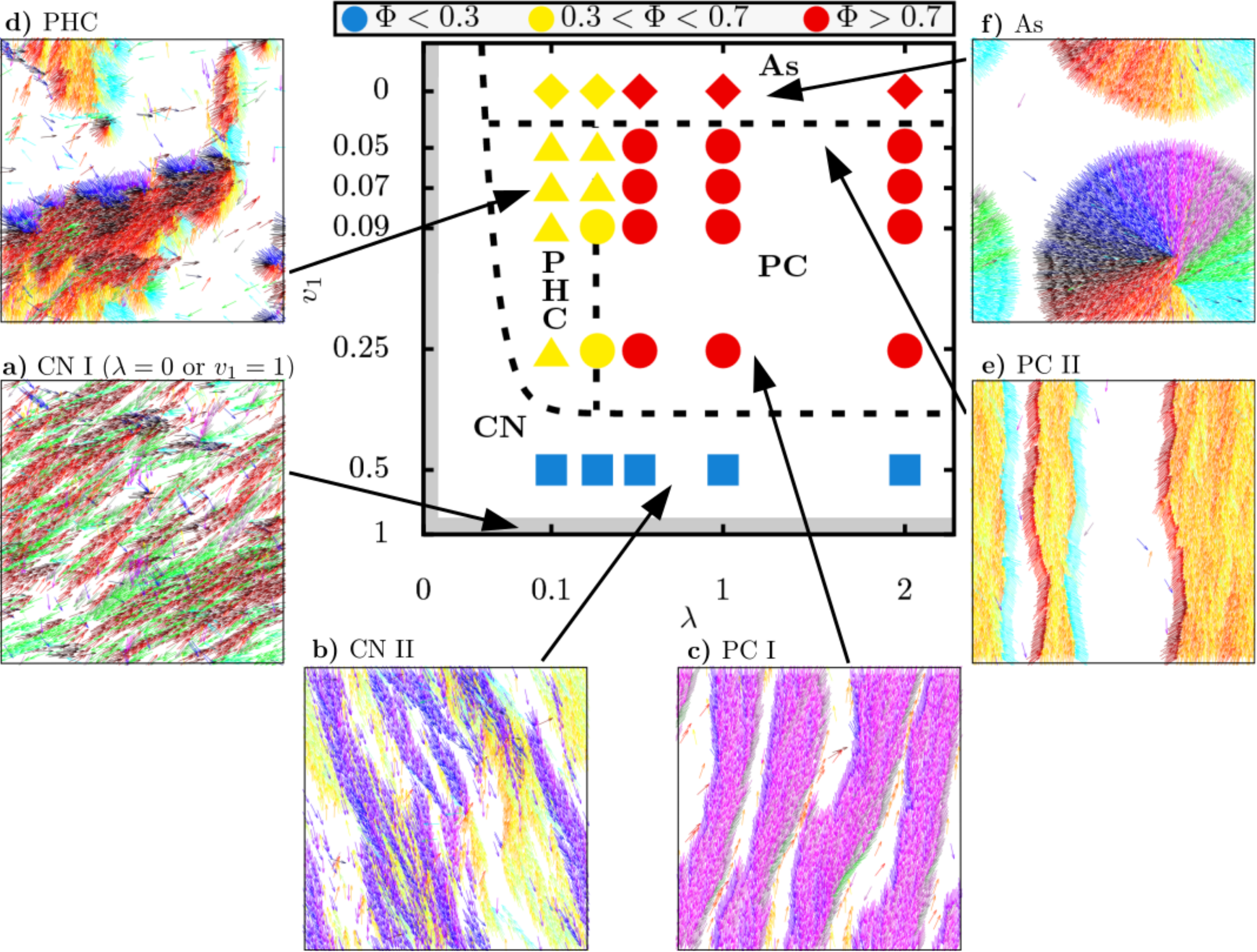}
    \caption{(Color online) Snapshots and phase diagram of SPR systems with aspect ratio $a/b=9$ $(n=18)$, ${\rho}_0{L}^2=12.8$, $E/k_\textrm{B}T=5$, and $\textrm{Pe}=400$. \CAV{\textbf{a)} Clustered nematic phase (CN I), system with constant propulsion, i.e., $\lambda=0$ or $v_1=1$. \textbf{b)} Clustered nematic phase (CN II), system with $\lambda=2$ and $v_1=0.5$. \textbf{c)} Polar clusters phase (PC I), system with $\lambda=2$ and $v_1=0.25$. \textbf{d)} Polar hedgehog clusters phase (PHC), system with $\lambda=0.4$ and $v_1=0.09$. \textbf{e)} Polar clusters with perpendicular rods at the borders phase (PC II), system with $\lambda=2$ and $v_1=0.09$. \textbf{f)} Asters phase (AS), system with  $\lambda=0.4$ and $v_1=0$.} \CAV{The gray regions in the phase diagram indicate systems with density-independent propulsion, which correspond with $\lambda=0$ or $v_1=1$. In these regions the rods form a clustered nematic phase (CN).} 
    \CAV{In the phase diagram squares represent clustered nematic rods (CN), triangles represent polar hedgehog clusters (PHC), circles represent polar clusters (PC), and diamonds represent asters (AS).} The points are colored according to the value of the polar order parameter $\Phi$, see the legend. Note that the base propulsion weight $v_1$ appears in the vertical axis top to bottom. \CAV{The penetrability coefficients range from $Q=0.8$ for density-independent systems ($\lambda=0$ or $v_1=1$) to $Q=0.04$ for systems with $\lambda=2$ and $v_1=0$.} Selected movies are presented in the Supplemental Material.}
    \label{snapshots_dens_l18} 
\end{figure*}
    
\section{PHASE DIAGRAMS AND ALIGNMENT MECHANISMS}\label{sec:results}
\subsection{PHASE BEHAVIOR}\label{sec:phase_diagrams}
Density-dependent reduced propulsion force introduces a rich variety of dynamical structures, depending on aspect ratio $a/b$, deceleration ratio $\lambda$, and weight of the base propulsion force $v_1$. 
\CAV{Figure~\ref{snapshots_dens_l9} shows a phase diagram and simulation snapshots for} rods with \CAV{aspect ratio $a/b=4.5$ (corresponding to $n=9$),} $E/k_\textrm{B}T=5$, and Pe$=100$.
For density-independent propulsion force \CAV{($\lambda=0$ or $v_1=1$)}, and for systems with \CAV{small $\lambda$ or large $v_1$}, the rods form a clustered nematic phase (CN). 
For intermediate and large $\lambda$ and intermediate and small $v_1$, the rods form clusters with domains (CD).
Finally, for intermediate and small $v_1$ and for small $\lambda$, the rods form round clusters (RC).
    
\CAV{Rods in the CN phase form small motile polar clusters, but the overall rod order is nematic. Systems with $\lambda>0$ (CN II) show larger clusters than systems with $\lambda=0$ or $v_1=1$ (CN I).}
In the CD phase, clusters are composed of large polar domains and span the entire system (CD I). \CAV{For systems with $v_1=0$ (CD II), rods both in bulk and at the borders are oriented perpendicular to the cluster borders.}
In the RC phase, we observe round clusters with small polar domains. Here, the rods at the cluster borders are perpendicularly oriented with respect to the borders.
    
\CAV{Figure~\ref{snapshots_dens_l18} shows a phase diagram and simulation snapshots for rods with aspect ratio $a/b=9$ (corresponding to $n=18$), $E/k_\textrm{B}T=5$ and Pe$=400$, $E/k_\textrm{B}T=5$, and Pe$=100$
\footnote{Rods with $a/b=9$), $\text{Pe}=100$ and $E/k_\textrm{B}T=5$ at the same density form polar clusters, see Fig.~S4 in the Supporting Information. For these systems, we do not observe a nematic-to-polar transition caused by the density-dependent propulsion, because the system with $\lambda=0$ or $v_1=1$ is already polar.}.}
For density-independent propulsion force ($\lambda=0$ or $v_1=1$), \CAV{and for small $\lambda$ or large $v_1$ the rods form a clustered nematic phase (CN).}
For \CAV{small } and intermediate $v_1$, the rods form polar clusters (PC).
For small $\lambda$ and intermediate and small $v_1$, the rods form polar hedgehog clusters (PHC).
Finally, for $v_1=0$, the rods form asters (AS). 
    
\begin{figure*}
    \centering
    \includegraphics[width=1.9\columnwidth]{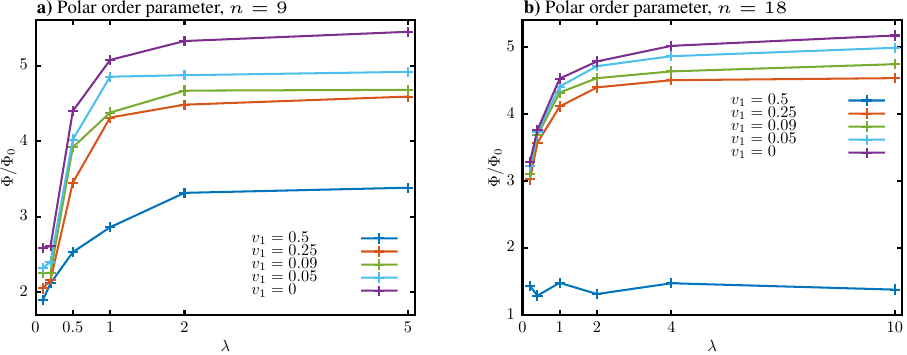} 
    \caption{(Color online) Polar order parameter $\Phi$ versus deceleration ratio $\lambda$ for various values of $v_1$. $\textbf{a)}$ Systems with aspect ratio $a/b=4.5$ $(n=9)$, ${\rho}_0{L}^2=6.4$, $E/k_\textrm{B}T=5$, and $\textrm{Pe}=100$. $\textbf{b)}$ Systems with aspect ratio $a/b=9$ $(n=18)$, ${\rho}_0{L}^2=12.8$, $E/k_\textrm{B}T=5$, and $\textrm{Pe}=400$. The polar order parameter for the respective density-independent propulsion force systems is $\Phi_0=0.074$ for short rods, and $\Phi_0=0.168$ for long rods.}
    \label{polar_order}
\end{figure*}
    
In the CN phase, rods form motile polar clusters but the overall orientation is nematic \CAV{as the short-rod systems. Similarly, for small and finite $\lambda$ (CN II) show larger clusters than systems with $\lambda=0$ or $v_1=1$ (CN I)}.
In the PC phase, \CAV{systems with small $v_1$ (PC I) have perpendicular rods at the cluster borders. The perpendicularity of the rods at the borders increases the cluster stability (PC II).}
In the PHC phase, we find large polar worm-like clusters with perpendicular rods at the borders. 
In the AS phase, rods form round clusters with large polar domains. Rods both within the cluster and at the border are perpendicular with respect to the aster border. 

\CAV{For both aspect ratios, the polar order parameter provides a good indication for the phase boundaries in Figs.~\ref{snapshots_dens_l9} and \ref{snapshots_dens_l18}. Enhanced polar order, enhanced cluster formation, perpendicularity at cluster borders, and nearly empty low-density regions induced by a density-dependent reduced propulsion are observed both for short and long rods. However, the differences in parameter space are more prominent for long rods that form more ordered structures than short rods.}

\CAV{This is related to the isotropic and nematic order in lyotropic liquid crystalline systems, where a minimum aspect ratio of about 5 is required for nematic order to appear \cite{frenkel_nematic_1990}.}
\CAV{Summarizing the discussion of the phase diagrams,} density-dependent self-propulsion enhances local polar alignment, see Sec.~\ref{sec:polar_order}. The alignment and cluster formation observed for classical SPRs ($\lambda=0$ or $v_1=1$) are caused by the rod-rod repulsive interaction, i.e., the rod energy barrier $E$. For $\lambda>0$, the density-dependent reduced propulsion allows the steric interaction to act longer, which enhances rod alignment. \CAV{Furthermore, density-dependent} reduced propulsion enhances clustering in two respects, see Sec.~\ref{sec:density_cluster}. On the one hand, there is a \CAV{higher probability} for the rods to align, because the rod energy barrier acts longer. 
On the other hand, \CAV{a decreased propulsion force favours rod} trapping. A rod that comes close to a cluster moves more slowly. This decreases the probability of a rod to exit the cluster. These two effects lead to the formation of loosely-packed CD phase. 
\CAV{Finally, the reduced propulsion at high densities effectively increases the friction as rods interact, which induces perpendicularity of rods at the cluster borders. A more detailed discussion} of this mechanism is given in Sec.~\ref{sec:asters}. The rod perpendicularity increases with increasing slowing-down. For systems with $v_1=0$, rods in clusters become effectively passive. \CAV{In these systems, we observe asters, as well as clusters with domains where not only rods at the borders, but all rods are oriented perpendicularly at the cluster surface.}
    
\subsection{DENSITY-DEPENDENT PROPULSION ENHANCES POLARITY}\label{sec:polar_order}
To quantify the increased rod alignment and polarity caused by the density-dependent self-propulsion we calculate the nematic order parameter
    \begin{equation}
    S = \left\langle\sum_{i\neq j}^{N} \frac{\cos{(2(\theta_i - \theta_j))}}{N(N-1)}\right\rangle
    \label{eqn:nematic_order}
    \end{equation}
    and the polar order parameter
    \begin{equation}
    \Phi = \left\langle\sum_{i\neq j}^{N} \frac{\cos{(\theta_i - \theta_j)}}{N(N-1)}\right\rangle \text{.}
    \label{eqn:polar_order}
    \end{equation}
The averages are taken over square cells of size $4{L}^2$. Here, $S=0$ corresponds to an isotropic state, $\Phi=0\text{ and }S=1$ to a nematic state, and $\Phi=1\text{ and }S=1$ to a polar state. The values for $S\text{ and }\Phi$ for various simulations are provided in the Supporting Information in Tabs.~S1 and S2. 
    
\CAV{The dependence of the polar order parameter $\Phi/\Phi_0$ on the density-dependent reduced propulsion is similar for rods with aspect ratios $a/b=4.5$ and 9 \footnote{For density-independent propulsion force, the polar order parameter $\Phi_0$, is higher for rods with 18 beads than for rods with 9 beads. We observe an increase in the polar order parameter as the energy barrier increases, see Figs.~S1 and S2 in the Supporting Information.}.} We observe an increase in the polar order parameter with decreasing $v_1$ \CAV{and with decreasing $\lambda$}, see Fig.~\ref{polar_order}. \CAV{The polar order parameter can be used as one criterion to determine phase boundaries and is therefore also represented by the color of the symbols in Figs.~\ref{snapshots_dens_l9} and \ref{snapshots_dens_l18}.
For $\lambda\lesssim1$, $\Phi$ increases with increasing $\lambda$, while for $\lambda\gtrsim1$, the polar order parameter remains roughly constant, see Fig.~\ref{polar_order}a.}
\CAV{For systems with $\lambda\gtrsim1$ and parallel rods, $\langle m\rangle/n\approx2.5$. Thus even for the smallest base propulsion weight $v_1=0$, $v_0\exp(-\lambda m)\approx0$ and $F_\text{p}\approx v_1F_0$.}
All systems show qualitatively similar behavior independent of the value of $v_1$, except for 18-bead rods in the aligned phase (Al, $v_1=0.5$), see Fig.~\ref{polar_order}b. \CAV{The sharp decrease in polar order compared with systems for smaller values of $v_1$ results from the formation of a lane phase that is also found for constant-propulsion rods.}
    
\subsection{DENSITY-DEPENDENT PROPULSION ENHANCES CLUSTERING}\label{sec:density_cluster}
\begin{figure*}
	\centering
    \includegraphics[width=2\columnwidth]{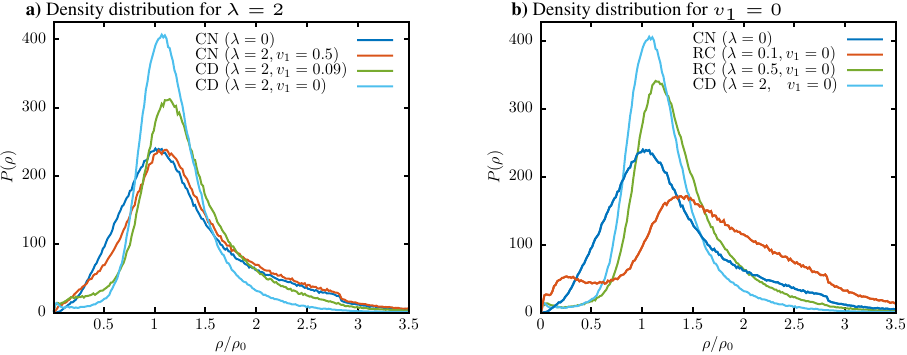} 
    \caption{(Color online) Density distribution $P(\rho)$ versus rod density $\rho$ for systems with aspect ratio $a/b=4.5$ $(n=9)$, ${\rho}_0{L}^2=6.4$, $E/k_\textrm{B}T=5$, $\textrm{Pe}=100$. \CAV{Clustered nematic phase (CN), clusters-with-domains phase (CD), round clusters (RC), values of $\lambda$ and $v_1$ are given in the legend.} \textbf{a)} Systems with \CAV{$\lambda=2$}, \textbf{b)} systems with $v_1=0$.}
    \label{density_l9}
\end{figure*}

\begin{figure*}
    \centering
    \includegraphics[width=2\columnwidth]{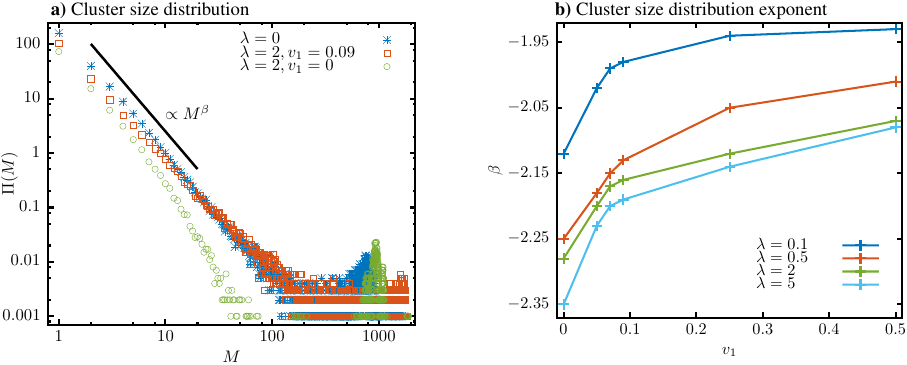} 
    \caption{(Color online) \textbf{a)} Cluster size distribution $\Pi(M)$ versus cluster size $M$ for systems with aspect ratio $a/b=4.5$ $(n=9)$, ${\rho}_0{L}^2=6.4$, $E/k_\textrm{B}T=5$, $\textrm{Pe}=100$. \CAV{Clustered nematic phase (CN), clusters-with-domains phase (CD), values of $\lambda$ and $v_1$ are given in the legend.} \textbf{b)} \CAV{Exponents $\beta$ of power-law fits to the cluster distributions for small $M$} versus the weight of the base propulsion, $v_1$, for various values of $\lambda$. \CAV{For the density-independent system ($\lambda=0$ or $v_1$=1), $\beta=-1.91$.}}
    \label{cluster_dist_l9}
\end{figure*}

We quantify the effect of density-dependent propulsion on the clustering process using rod-density distributions $P(\rho)$ and cluster size distributions $\Pi(M)$, see Figs.~\ref{density_l9} and \ref{cluster_dist_l9}. The local rod densities are calculated using Voronoi tessellation as the inverse of the area of the Voronoi cells, where the centers-of-mass of the rods are used as the tessellation seeds. 
Because many of our systems have very few rods in the low-density regions, the density distributions are, generally, single-peaked functions, see Fig.~\ref{density_l9}. \footnote{A peak for $\rho=0$ is lacking in our density distributions, because when using the Voronoi tessellation the density distribution is weighted by rod the density itself.}
For the cluster-size distributions, we consider two rods to be in the same cluster if their nearest distance is less than $2r_\text{cut}$ and if their orientations differ by less than $15\textordmasculine$ \cite{abkenar_collective_2013}.

Rods in the CN phase form clusters of different size without a clear spatial separation of high and low-density regions, see Fig.~\ref{snapshots_dens_l9}a. Therefore, $P(\rho)$ shows a single but broad peak that is positively skewed, see Fig.~\ref{density_l9}a. The peak at $\rho/\rho_0=1.05$ is only slightly above the average density $\rho_0$. The peak position is independent of $v_1$, while the peak height increases with decreasing $v_1$.  
The density distributions for the \CAV{CD phase, however,} are significantly narrower. For these systems with small $v_1$, the probability to find \CAV{densities} $\rho/\rho_0=0.5$ is \CAV{very small, $\rho/\rho_0=0.5$ corresponds to the rod density obtained from the Voronoi cells for rods at the cluster borders. 
A strong density dependence of the} reduced propulsion thus reduces the probability for rods to leave the cluster.
\CAV{In particular, rod perpendicularity at the borders leads} to the formation of clusters with very sharp interfaces, \CAV{compare Figs.~\ref{snapshots_dens_l9} and \ref{snapshots_dens_l18}.}
    
For systems with $v_1=0$, the height of the peak of the density distribution \CAV{does not increase monotonically} with increasing $\lambda$, see Fig.~\ref{density_l9}b. For the RC system \CAV{with $\lambda$=0.1, the height of the peak is lower than for the CN phase with $\lambda=0$ or $v_1=1$. For $\lambda \geq 0.5$, the peak height increases with increasing $\lambda$ and is higher than for the CN phase.}
The density distribution for the RC phase \CAV{with $\lambda=0.1$} is bimodal, where the first peak at $\rho/\rho_0\approx0.25$ corresponds to rods in the low-density region. Here, the low-density region is more populated than in the other phases, see Fig.~\ref{snapshots_dens_l9}. 
The second peak of the density distribution at $\rho/\rho_0^2\approx1.4$ corresponds to rods in the cluster. 
For \CAV{the RC system with $\lambda=0.5$, the system is at the border of the RC and CD phases. Therefore} $P(\rho)$ has a single broad peak at $\rho/\rho_0\approx1.2$, where rods in the high-density region are more loosely packed than in the RC phase.
For \CAV{the CD system with $\lambda=2$}, the peak is approximately at the same position as for the density-independent system, $\rho/\rho_0\approx1.1$. 
\CAV{For systems with $\lambda\leq0.5$, we find} very sharp interfaces so $P(\rho/\rho_0\leq0.5)$ is very small, as discussed above. 
Density distributions for rods with aspect ratio $a/b=9$ are shown in the Supporting Information. The \CAV{position of the peak for} density distributions functions for long \CAV{rods is shifted to} $\rho/\rho_0\approx1.2$.
   
    
Cluster-size distributions for rods with 9 beads are shown in Fig.~\ref{cluster_dist_l9}a. \CAV{Although the distributions do not always show true power laws for several orders of magnitude, they can be characterized by effective power law $\Pi(M)\propto M^\beta$ in the limit of small cluster sizes, $M \le 15$.}
As the density-dependence of the reduced propulsion becomes more pronounced, the number of small clusters decreases and the rods form "loose" system-spanning clusters as $\lambda$ increases and $v_1$ decreases. 
The cluster size distributions for the CN system with $\lambda=0$ or $v_1=1$ and for the CD system with $\lambda=2$, $v_1=0.09$ are very similar. However, $\Pi(M)$ drastically changes for the CD system with $\lambda=2$, $v_1=0$.
\CAV{All cluster-size distributions show a peak at $M\simeq1000$, which corresponds to system-spanning clusters that consist of a large number of rods. In particular, for systems with perpendicular rods at the border, $\Pi(M)$ is smaller and decreases faster with increasing $M$ for small cluster sizes.}
    
The exponent $\beta$ of the cluster-size distribution decreases with increasing $\lambda$ and decreasing $v_1$, see Fig.~\ref{cluster_dist_l9}b
\CAV{\footnote{Systems with small $v_1$, $\Pi(M)$ show a first power-law decay for $\leq15$ and a second power-law decay with a more negative exponent for $15 < M\leq100$, see CD system with $\lambda=2$, $v_1=0$. The exponents $\beta$ shown in Fig.~\ref{cluster_dist_l9} have been calculated for the power-law decay at small cluster sizes.}.
$\beta$ decays slowly with $v_1$ for $v_1\leq0.09$, which corresponds to the CN and CD phases. However, $\beta$ decreases sharply for $v_1\leq0.09$, which corresponds to CD and RC phases. The decrease of $\beta$ with decreasing $v_1$ is caused by the perpendicularly-oriented rods at the cluster borders. Because of the hindered rotational diffusion of SPRs that newly join clusters, the strong depletion of rods in the low-density region disfavors the formation of many small clusters.}

    \begin{figure*}
        \centering
        \includegraphics[width=2\columnwidth]{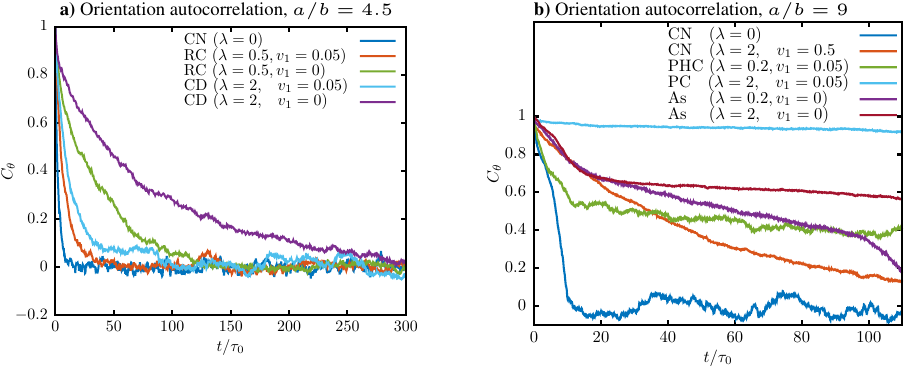} 
        \caption{(Color online) Rod orientation autocorrelation function $C_\theta$ versus lag time $t$. \CAV{Clustered nematic phase (CN), clusters-with-domains phase (CD), round clusters (RC), polar hedgehog clusters (PHC), polar clusters (PC), asters (AS), values of $\lambda$ and $v_1$ are given in the legend.} \textbf{a)} Systems with aspect ratio $a/b=4.5$ $(n=9)$, ${\rho}_0{L}^2=6.4$, $E/k_\textrm{B}T=5$, $\textrm{Pe}=100$, for various values of $\lambda\text{ and }v_1$. \textbf{b)} Systems with aspect ratio $a/b=9$ $(n=18)$, ${\rho}_0{L}^2=12.8$, $E/k_\textrm{B}T=5$, $\textrm{Pe}=400$, for various values of $\lambda\text{ and }v_1$.}
        \label{orientation_correl}
    \end{figure*}
    \begin{figure}
        \centering
        \includegraphics[width=0.92\columnwidth]{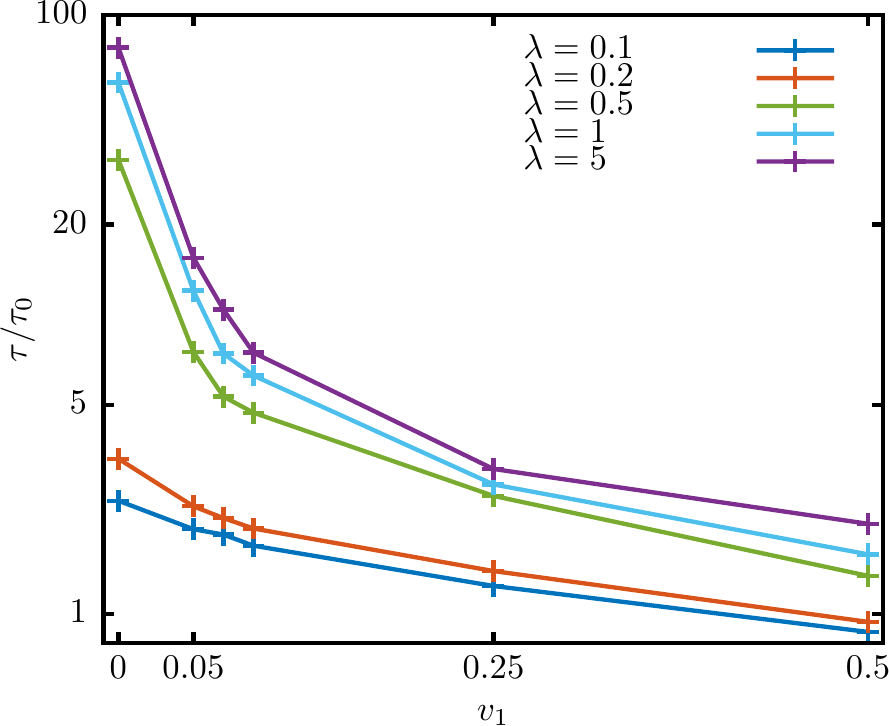} 
        \caption{(Color online) Rod orientation autocorrelation times $\tau$ versus the weight of the base propulsion $v_1$. Systems with aspect ratio $a/b=4.5$ $(n=9)$, ${\rho}_0{L}^2=6.4$, $E/k_\textrm{B}T=5$, $\textrm{Pe}=100$, for various values of $\lambda$. \CAV{For the density-independent system ($\lambda=0$ or $v_1$=1), $\tau=2.21\tau_0$.}}
        \label{orientation_decorrel_l9}
    \end{figure}
    
\subsection{DENSITY-DEPENDENT PROPULSION INDUCES PERPENDICULARITY}\label{sec:asters}
For systems with density-independent propulsion ($\lambda=0$ or $v_1=1$), the rods form elongated clusters. In contrast, for systems with density-dependent reduced propulsion ($\lambda>0$), we find perpendicular rods at the borders of the clusters: RC, CD with perpendicular borders, PHC, PC phases with perpendicular rods at the borders, and AS phases, see Figs.~\ref{snapshots_dens_l9} and \ref{snapshots_dens_l18}. 
For rods with small $v_1$, the propulsion force strongly decreases as soon as a rod comes in contact with a cluster. Therefore, rods do not easily slide along a \CAV{cluster borders and alignment due to the propulsion-induced torque is also strongly decreased. This enhances the probability for a rod to meet another rod that slides along the border in opposite orientation. Density-dependent reduced propulsion thus induces jamming between rods. In addition, jammed structures easily trap other rods that newly join a cluster, giving rise to small hedgehog-like clusters with perpendicular rods at the borders.} Once the cluster border is filled with rods, the rods align perpendicularly rather than forming separate hedgehog-like aggregates, to maximize packing. A figure detailing the formation PHC, PC phases with perpendicular rods, and AS phases can be found in the SI. 

\CAV{For polar hedgehog clusters and polar clusters with perpendicular rods at the borders ($v_1>0$), the rods in the center of the cluster are still propelled, which leads to rod alignment due to the constant-propulsion SPR alignment mechanism. Therefore, systems with $v_1>0$ are overall polar. 
For systems with $v_1=0$ and sufficiently high $\lambda$, the rods become passive as soon as they collide with other rods. Therefore, the local structure in the interior of the cluster is determined only by steric alignment and optimal packing throughout the cluster.
There is no propulsion-induced alignment, all rods are perpendicular with respect to the borders, forming AS and CD phases that are typically not observed for constant-propulsion SPRs. RC phases have earlier been reported also for very rough SPRs that interlock \cite{yang_swarm_2010}.}

\begin{figure*}
    \centering
    \includegraphics[width=\textwidth]{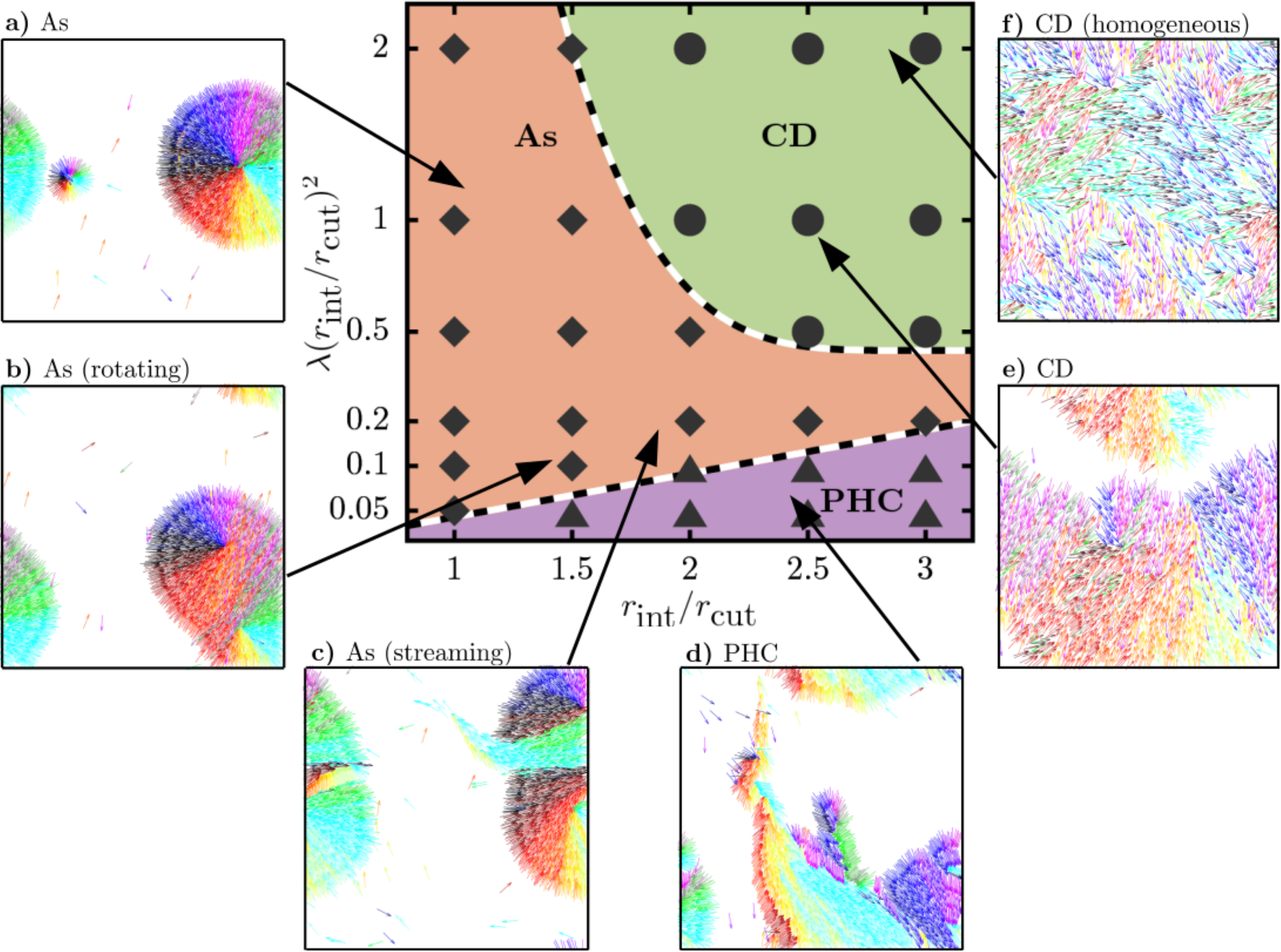}
    \caption{\CAV{(Color online) Phase diagram for various $r_\textrm{int}$ and $\lambda_1$, and simulation snapshots of SPR systems with aspect ratio $a/b=9$, $\rho_0 L^2=6.4$, $E/k_\textrm{B}T=5$, $\textrm{Pe}=400$, and $v_1=0$. \textbf{a)} Asters (AS), system with $\lambda_1=0.1$ and $r_\textrm{int}/r_\textrm{cut}=1$. \textbf{b)} Rotating aster, system with $\lambda_1=0.1$ and $r_\textrm{int}/r_\textrm{cut}=1.5$. \textbf{c)} Streaming jet aster, system with $\lambda_1=0.2$ and $r_\textrm{int}/r_\textrm{cut}=2$. \textbf{d)} Polar hedgehog clusters (PHC), system with $\lambda_1=0.1$ and $r_\textrm{int}/r_\textrm{cut}=3$. \textbf{e)} Clusters with domains (CD), system with $\lambda_1=1$ and $r_\textrm{int}/r_\textrm{cut}=2.5$. \textbf{f)} Clusters with domains with homogeneous density, system with $\lambda_1=2$ and $r_\textrm{int}/r_\textrm{cut}=3$. In the phase diagram, triangles represent polar hedgehog clusters (PHC), diamonds represent asters (AS), and circles represent clusters with domains (CD). Selected movies are presented in the Supplemental Material.} }
    \label{quorum} 
\end{figure*}
    \begin{figure}
        \centering
        \includegraphics[width=0.92\columnwidth]{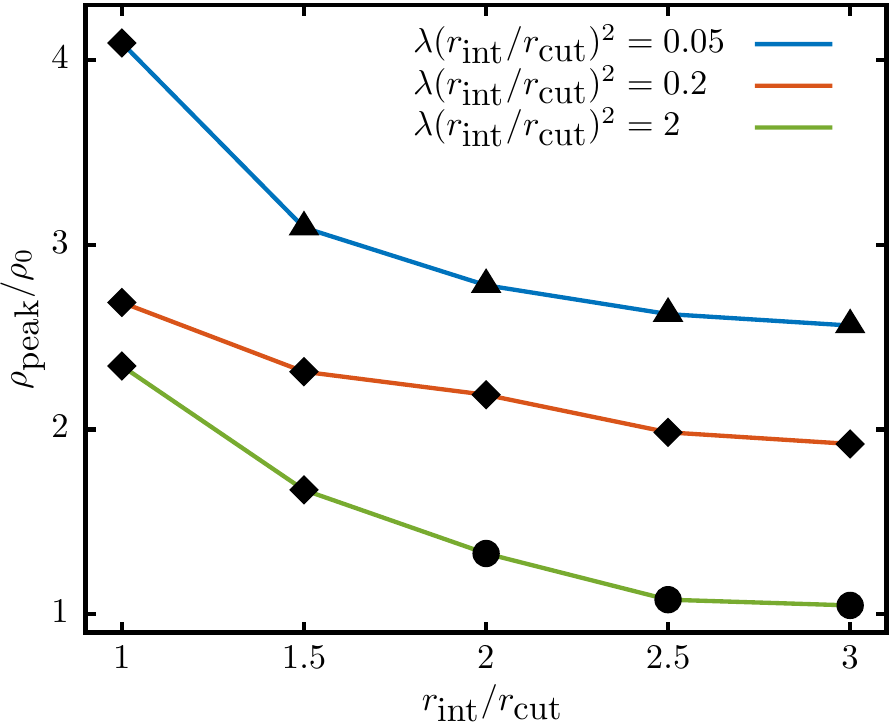} 
        \caption{\CAV{(Color online) Position of the density distribution peak, $\rho_\textrm{peak}$, versus quorum sensing interaction radius $r_\textrm{int}$ for systems with aspect ratio $a/b=9$ $(n=18)$, ${\rho}_0{L}^2=6.4$, $E/k_\textrm{B}T=5$, $\textrm{Pe}=400$, for various values of $\lambda_1$. Triangles represent a polar-hedgehog-cluster phase (PHC), diamonds an asters phase (AS), and circles a clusters-with-domains phase (CD)}}
        \label{dens_quorum}
    \end{figure}

\subsection{ROD DYNAMICS}\label{sec:orient_correl}
So far we have only described the structure of clusters and interfaces. \CAV{To study rod dynamics in steady state more systematically,} we calculate the rod orientation autocorrelation function
    \begin{equation}
    \label{eqn:orient_correl}
    C_\theta(t)=\left\langle \textbf{l}_i(t'+t)\cdot\textbf{l}_i(t') \right\rangle \textrm{,}
    \end{equation} 
where $\textbf{l}_i(t')$ is the orientation vector of rod $i$ at time $t'$, and $t$ is the lag time.
For rods with aspect ratio $a/b=4.5$, the rod orientation autocorrelation function decreases exponentially $C_\theta=e^{-t/\tau}$ where $\tau$ is the relaxation time, see Fig.~\ref{orientation_correl}a. $C_\theta$ decorrelates more slowly as $\lambda$ increases and $v_1$ decreases. But $v_1$ has a stronger effect in the reduced propulsion of the autocorrelation function than $\lambda$, compare the \CAV{dynamics for systems with $v_1=0.05$ and for systems with $v_1=0$.}
For systems of rods with aspect ratio $a/b=9$, $C_\theta$ is not always an exponentially decreasing function, see Fig.~\ref{orientation_correl}b. The functional form of the $C_\theta$ strongly depends on the structure formed by the rods. For the CN phase \CAV{with $\lambda=0$ or $v_1=1$}, the rod orientation quickly becomes uncorrelated. For systems with density-dependent reduced propulsion, $C_\theta$ decorrelates more slowly. Density-dependent reduced propulsion slows down rod dynamics, and thus increases the autocorrelation time for rod orientation.

For rods with aspect ratio $a/b=9$, $C_\theta(t)$ does not always decrease exponentially.
In PC systems with perpendicular rods at the borders, \CAV{system with $\lambda=2$, $v_1=0.05$}, the perpendicular rods at the border stabilize the polar clusters and $C_\theta$ roughly remains constant.
For \CAV{the CN system with} $\lambda=2$, $v_1=0.5$, $C_\theta$ decorrelates slowly. For the \CAV{PHC system with} $\lambda=0.2$, $v_1=0.05$, $C_\theta$ decorrelates quickly to a finite value, $C_\theta(t\gg\tau_0)\backsimeq0.4$. Polar hedgehog clusters are dynamic at short times, rods can wiggle, but the perpendicular rods at the cluster borders stabilize the overall orientation. 
For the AS system with $\lambda=0.2$, $C_\theta$ shows three linear regimes. A first regime with a fast decay, corresponding to single-rod dynamics in the dense region; a second slow decay, corresponding to rod collective behavior inside the aster; a third fast decay, corresponding to long-time behavior of single rods that exit the aster. 
For the AS system with $\lambda=2$, $v_1=0$, very few rods escape the asters. This leads to two decay regimes: a first quick decay, corresponding to single-rod dynamics in the dense region, and a second slow decay, corresponding to collective rod dynamics inside the aster.
    
For rods with aspect ratio $a/b=4.5$, the relaxation time $\tau$ increases with increasing $\lambda$ and with decreasing $v_1$, see Fig.~\ref{orientation_decorrel_l9}. For \CAV{systems with $\lambda\leq0.2$,} $\tau$ decreases at a constant rate because the rod moves faster through an RC or a CN phase with increasing $v_1$. Rod dynamics in both phases is therefore very similar. For \CAV{systems with $\lambda>0.2$} we find two different decays.
For $v_1\leq0.09$,
$\tau$ decreases sharply, whereas for $v_1>0.09$,  $\tau$ decreases slowly. \CAV{The highly-ordered CD phase with perpendicular rods at the border for small $v_1$ transitions to a considerably more disordered CN phase for large $v_1$. Therefore, the rotational autocorrelation time decreases sharply at the phase boundary.}

\section{ROLE OF THE ``QUORUM-SENSING'' INTERACTION RADIUS}\label{sec:quorum}
\CAV{An important feature of the quorum sensing mechanism is that the range of this interaction can be chosen independently from the steric repulsion range. This strongly affects the emerging structures. As the range of the quorum sensing interaction increases, the rods become more homogeneously distributed and rod perpendicularity decreases. Simulation results are shown in Fig.~\ref{quorum}, which characterizes dynamics and collective behavior of rods with $a/b=9$, $\rho_0L^2=6.4$, and $v_1=0$, as a function of interaction radius $r_{\textrm{int}}$ and deceleration ratio $\lambda_1$. 
For small $\lambda_1$, with increasing $r_{\rm int}$ asters first become more dynamic, intermittently ejecting streams of clustered rods, and finally polar hedgehog clusters develop. As $\lambda_1$ increases, rod dynamics slows down and the systems become more static. For large $r_{\rm int}$ and $\lambda_1$, we find a less dense CD phase. A particular case is the homogeneous-density system with $\lambda_1=2$ and $r_{\textrm{int}}/r_{\textrm{cut}}=3$.

The $\lambda_1$-$r_{\rm int}$ cut through the parameter space systematically characterizes the interplay of rod shape and chemical signaling. For the small interaction radius discussed in Sec.~\ref{sec:results}, shape plays the major role for determining the structure. With increasing $r_{\rm int}$ and increasing $\lambda_1$, chemical signaling becomes dominant. Static and highly symmetric round asters are stable when the steric interactions dominate. Systems with bacteria or phoretic particles that sense each other before they physically touch are thus less dense and less ordered. Interestingly, we find slightly elongated, circling asters close to the border between the AS and the PHC phase, e.g.\ for $\lambda_1\leq0.1$ and $r_\textrm{int}\geq2$. These dynamic asters are held together by a line tension induced by perpendicular rods at the aster border that exert an inward force, similar to molecules at interfaces in fluid-gas systems. Stresses are relaxed via intermittent jets of rods, similar to the mechanism in Ref.~\cite{peruani_stresses_2015}.

Figure~\ref{dens_quorum} shows the dependence of the position of the peak in the density distribution, $\rho_\textrm{peak}$, on the quorum sensing interaction radius $r_{\rm int}$. The peak position shifts to lower densities with both increasing $r_\textrm{int}$ and $\lambda_1$. For constant $\lambda_1$, the AS phase is denser than the PHC or the CD phase that occur at higher interaction radii. 
For large interaction radii, the SPRs already slow down before they touch. Therefore, the rods are effectively passive and motility-induced alignment becomes irrelevant. For $r_{\rm int}/r_{\rm cut} \geq 2.5$, the rods are homogeneously distributed with density $\rho_\textrm{peak} \approx \rho_0$, see simulation snapshot for the homogeneous CD phase in Fig.~\ref{quorum}.

}
    
\section{SUMMARY AND CONCLUSION}\label{sec:conclusions}
We have studied SPRs with density-dependent propulsion in quasi-two-dimensional systems with periodic boundary conditions. The rods interact via a capped repulsive potential, which mimics self-avoiding rods in a thin, three-dimensional slab, such that the rods have a finite probability to cross each other. We find a very rich collective behavior including several phases that are not observed for constant propulsion. \CAV{In general, reduced propulsion at high rod densities enhances polar order and cluster formation, and induces perpendicular orientation of rods at the cluster borders.}

\CAV{Rod-density distributions are an important observable to characterize phases and phase behavior of SPR systems. For constant-propulsion SPRs, phase separation is observed because of excluded volume interactions \cite{farrell_pattern_2012,peruani_mips_2015}. The rod-density distributions show two well-separated peaks, with the first peak representing the low-density region and the second peak the high-density region, respectively. On the contrary, the rod-density distributions for systems with quorum sensing--using Voronoi tessellation to calculate local rod densities--often show only one peak that corresponds to the high-density region. Low-density regions are almost devoid of rods due to the perpendicular orientation of the rods at the boundaries.  We thus find a positive feedback between density-dependent reduced propulsion and cluster growth.

Some previous studies have investigated various aspects of enhanced cluster formation due to MIPS in self-propelled spheres and discs \cite{cates_mips_2015}, as well as for systems that combine point particles that interact via the Vicsek model and quorum sensing \cite{farrell_pattern_2012,peruani_mips_2015}.} SPPs with Vicsek-type interactions and density-dependent reduced propulsion show similar dynamic phases to the ones observed in our systems: stripy, aster, moving clumps, and lane phases.       
\CAV{However, the models studied in Refs.~\cite{farrell_pattern_2012,peruani_mips_2015} do not allow a distinction between volume exclusion and quorum sensing as mechanisms of slowing down.}
Neither perpendicularity of the propulsion force at the border nor very regular asters  are observed, therefore low-density regions in systems with point particles are more populated than in our simulations.

\CAV{Density-dependent reduced propulsion introduces perpendicularity of the rods at the cluster borders. If the rods in the center of a cluster are still propelled, polar clusters form with perpendicular rods at the borders. If the propulsion force of the majority of rods in the cluster vanishes, stripe-like and aster-like clusters form, where all rods are oriented perpendicularly to the border. Increasing the interaction range used to calculate the density-dependent propulsion force, we can transition from a system, where only nearest neighbors are considered and steric interaction are important, to a system where the SPRs only interact via quorum sensing. The highest peak of the density distribution shifts to lower density with increasing interaction radius because the SPRs slow down before they touch. In particular for small $\lambda$, increased interaction radius implies that the asters become more dynamic and finally disintegrate into polar hedgehog clusters. The dynamic asters rotate and intermittently eject streams of rod clusters. A somewhat similar behavior has been observed for SPRs with constant propulsion in two dimensions, where very large} aster-like clusters eject streams of rods to relieve stresses \cite{peruani_stresses_2015}.

\CAV{In nature, an increased propulsion velocity with increased density has been observed in Bacillus subtilis populations \cite{sokolov_concentration_2007}. This mechanism has been found in computer simulations with Vicsek interactions to lead to band formation \cite{ohta_bands_2014,marchetti_accelerate_2010}.
In future, we therefore plan to employ our model to} study the collective behavior of SPRs with density-dependent enhanced propulsion force.
    
\acknowledgements
We thank Clemens Bechinger (Konstanz) for stimulating discussions. C.A.V. acknowledges support by the International Helmholtz Research School of Biophysics and Soft Matter (IHRS BioSoft). CPU time allowance from the J\"ulich Supercomputing Centre (JSC) is gratefully acknowledged.

\bibliography{reference}
    
\end{document}